\documentclass[prb,fleqn,12pt,showpacs]{revtex4-1}

\usepackage[colorlinks,linkcolor=blue,citecolor=blue,urlcolor=magenta]{hyperref}
\usepackage{graphicx}
\usepackage{bm,amsfonts,amsmath}
\usepackage{color,soul}

\begin{document}

\title{Magnetic excitations in frustrated fcc type-III antiferromagnet MnS$_2$}

\author{T. Chatterji}

\email[Email of corresponding author: ]{chatterji@ill.fr}

\affiliation{Institut Laue-Langevin, 71 Avenue des Martyrs, 38000 Grenoble, France}

\author{L.P. Regnault}

\affiliation{Institut Laue-Langevin, 71 Avenue des Martyrs, 38000 Grenoble, France}

\author{S. Ghosh}

\affiliation{Institute for Theoretical Physics, Technische Universit\"{a}t Dresden, 01062 Dresden, Germany}

\author{A. Singh}

\affiliation{Department of Physics, Indian Institute of Technology, Kanpur - 208016, India}
\date{\today}

\begin{abstract}

Spin wave dispersion in the frustrated fcc type-III antiferromagnet MnS$_2$ has been determined by inelastic neutron scattering using a triple-axis spectrometer. Existence of multiple spin wave branches, with significant separation between high-energy and low-energy modes highlighting the intrinsic magnetic frustration effect on the fcc lattice, is explained in terms of a spin wave analysis carried out for the antiferromagnetic Heisenberg model for this $S=5/2$ system with nearest and next-nearest-neighbor exchange interactions. Comparison of the calculated dispersion with spin wave measurement also reveals small suppression of magnetic frustration resulting from reduced exchange interaction between frustrated spins, possibly arising from anisotropic deformation of the cubic structure.

\end{abstract}


\maketitle 

\section{Introduction}

Frustrated magnetic systems are of considerable recent interest due to possibility of exotic magnetic states such as like spin-ice or spin-liquid states, with intensive research activities in novel systems such as the Compass-Heisenberg model on square lattice, Kitaev model on honeycomb lattice, pyrochlore lattice etc. While the relatively \lq{older}\rq Kagome and triangular lattice systems have been widely studied, frustration in the face-centered-cubic (fcc) lattice has attracted far less attention. Magnetic order on the fcc lattice is realized with three different kinds of antiferromagnetic (AF) arrangement of spins, distinguished by their alignments along the crystallographic $z$-axis. 

Antiferromagnetism on the fcc lattice was studied long ago using spin wave method \cite{Lines1963} and random phase approximation \cite{Lines1965a,Lines1965b} within spin models. Selection of collinear ground state by thermal fluctuation through the \lq{order by disorder}\rq~ effect was argued \cite{Henley1987}. Thermal fluctuation was also suggested to give rise to first-order magnetic transition in fcc lattice in renormalization group study \cite{Brazovskii1976}. More recently, Monte Carlo \cite{Gvozdikova2005} and first-principle methods \cite{Khmelevskyi2012} were employed to investigate AF order in fcc lattice. Within the itinerant electron approach, ground-state magnetic phase diagram and related metal-insulator transition was investigated using the slave-boson method \cite{Timirgazin2016}. Very recently, frustration effects on spin waves and magnetic instabilities were studied within the $t-t'$ Hubbard model \cite{Singh2017}. However, a detailed study of spin wave dispersion and comparison with experiments can be of particular interest in view of the measured spin wave dispersion in MnS$_2$ obtained from inelastic neutron scattering studies \cite{Chattopadhyay1989}. 

\begin{figure}
\includegraphics[width=55mm]{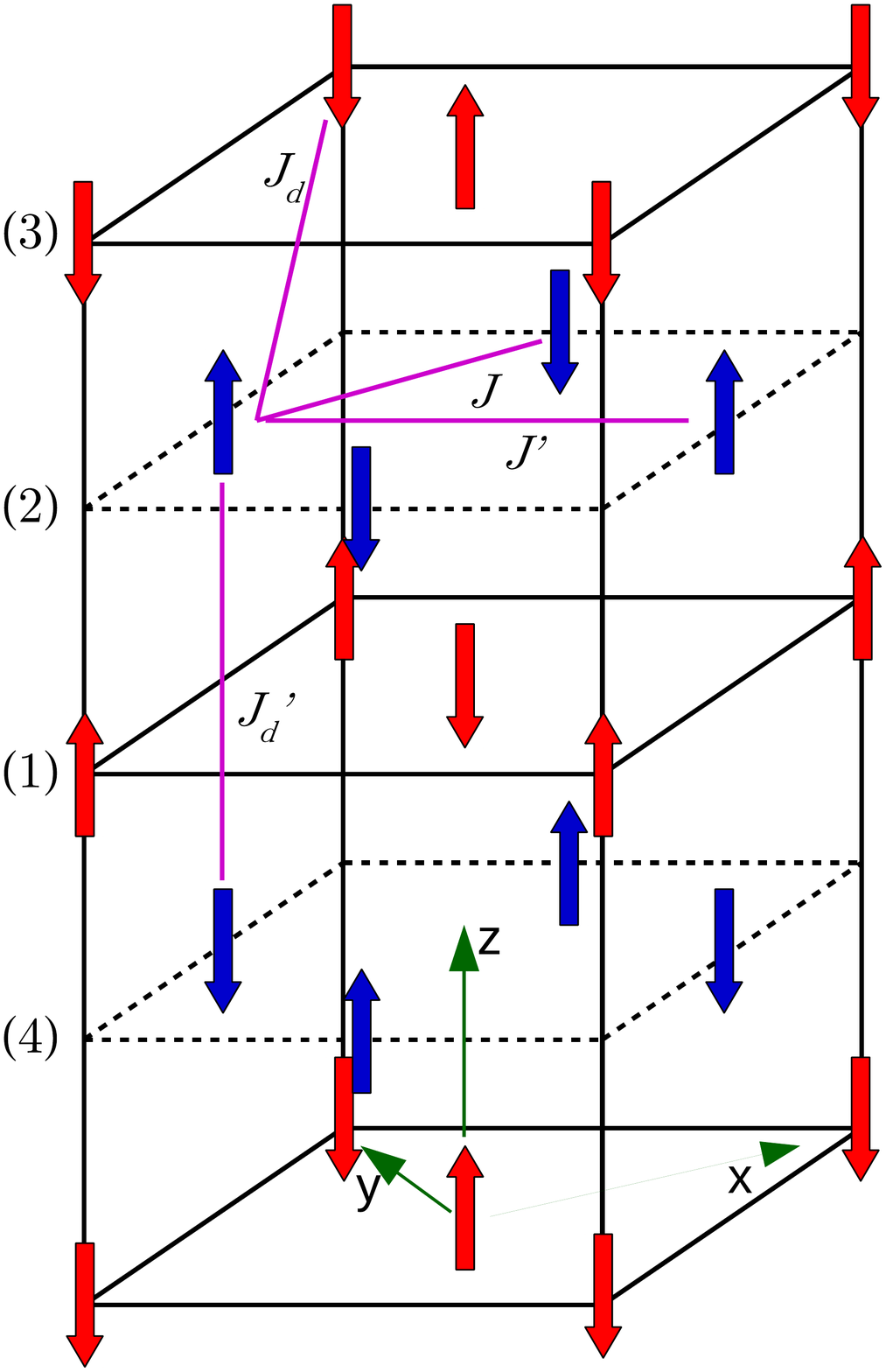}
\caption{Type-III antiferromagnetic order on the fcc lattice.  Planes shown in solid and dashed lines with spins in red and blue indicate the two identical fcc sublattices. The layers along the $z$ direction in the sequence $\alpha \alpha' \beta \beta'$... (labeled as 12341...) have antiferromagnetic arrangement of spins in the $x$ and $y$ directions, with two magnetic sublattices in each layer. The exchange constants for NN and NNN spins are $J,J'$ in the same layer, and $J_d,J_d '$ for different layers.}
\label{spin}
\end{figure}

\begin{figure}
\includegraphics[width=80mm]{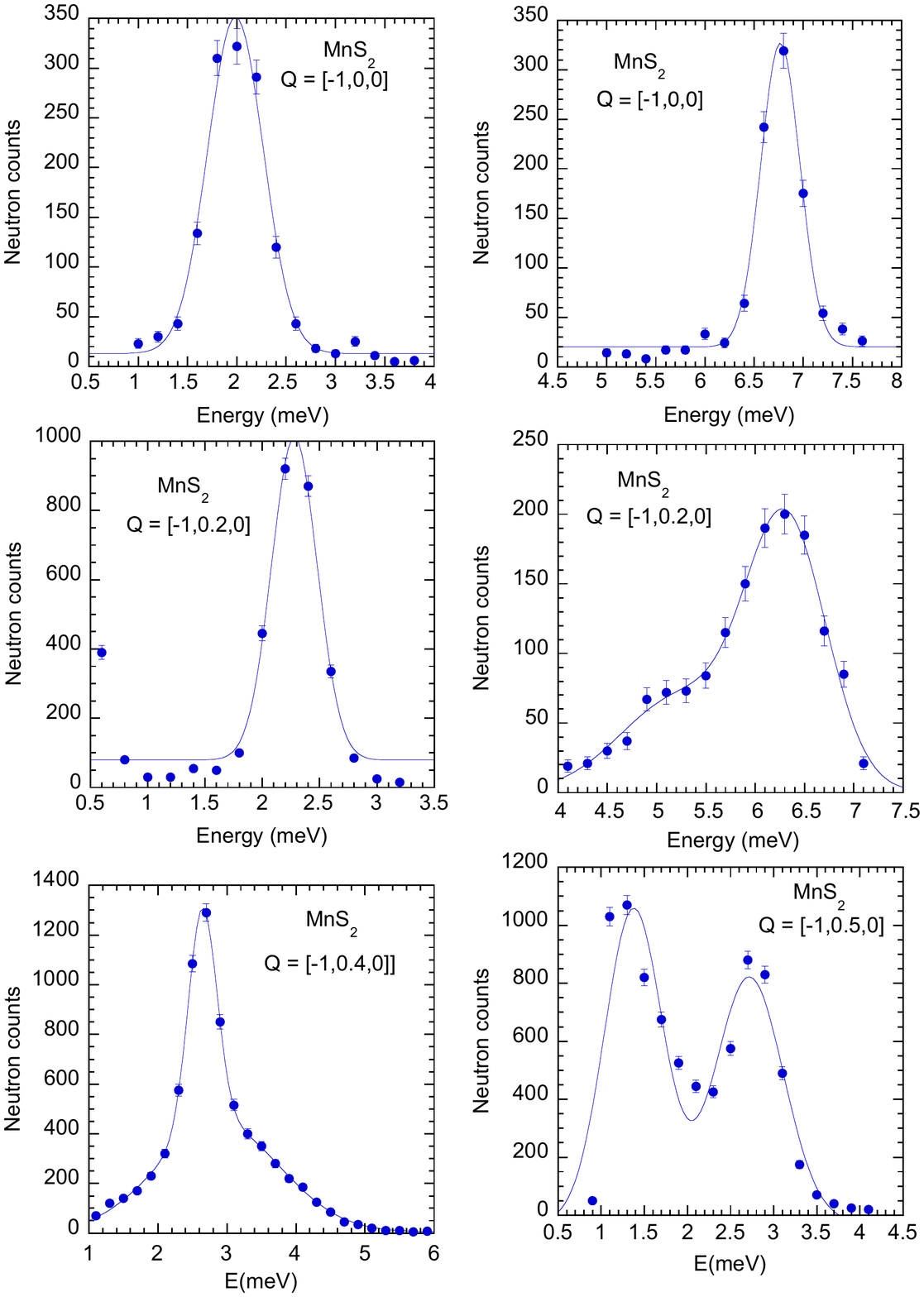}
\includegraphics[width=80mm]{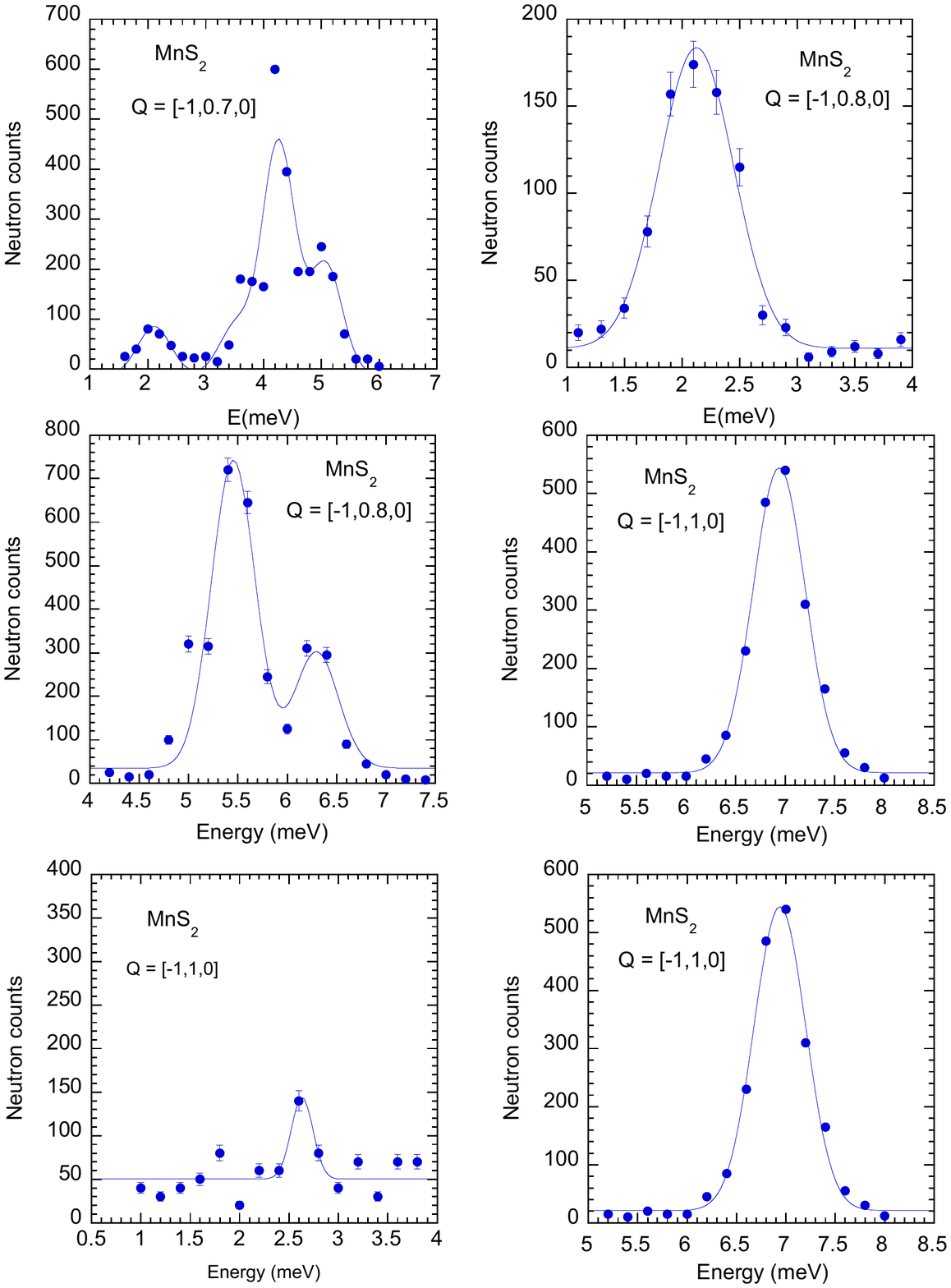}
\caption{Constant-$Q$ scans from MnS$_2$ that have been used to generate the spin wave dispersion along $[0,q_y,0]$.}
\label{scans}
\end{figure}

Magnetic order on the fcc lattice is manifested in three different kinds of spin alignments ranging from type-III and type-I in the 1:2 compounds MnS$_2$ and MnT$_2$ to type-II in the 1:1 compound MnO. All of them show a planar ($xy$ plane in Fig. \ref{spin}) antiferromagnetic (AF) arrangement of nearest-neighbor (NN) spins. But the order in the perpendicular ($z$) direction involving next-nearest-neighbor (NNN) spins distinguishes the three, with AF inter-layer order being realized for MnS$_2$ which is called type-III. For MnS$_2$, the magnetic phase transition is of first order occurring at $T_N$ = 48 K \cite{Hastings1976,Chattopadhyay1984,Westrum1970} with accompanying pseudo-tetragonal distortion \cite{Kimber2015}. 

As reported in a detailed crystal structure study \cite{Chattopadhyay1992}, $\rm MnS_2$ crystallizes with the pyrite structure with the primitive cubic space group Pa-3 having no four-fold symmetry axis. The structure consists of $\rm Mn^{2+}$ cations and $\rm S_2 ^{2-}$ anions. The magnetic $\rm Mn^{2+}$ cations are in the $3d^5$ $^6 S_{5/2}$ state with no orbital moment. However, the magnetic $\rm Mn^{2+}$ ions occupy a fcc lattice if one ignores the presence of the non-magnetic $\rm S_2 ^{2-}$ anions.

A comprehensive understanding of the magnetic phase in MnS$_2$ requires understanding of not only the magnetic ground state, but magnetic excitation spectra as well. Therefore, we have performed inelastic neutron scattering (INS) investigations of the spin wave excitations in MnS$_2$ using a triple-axis spectrometer. The preliminary experiments were done long time ago but the results were only available in short paper in the conference proceedings \cite{Chattopadhyay1989}. In the present paper we wish to describe the experiments in more details than was done before. We present a suitable  Heisenberg spin model that is used for interpretation of the experimental results.
 
\section{Experimental methods}

A large natural single crystal (hauerite) of MnS$_2$ was used for the experiment. The crystal volume was about 1 cm$^3$. We performed inelastic neutron scattering experiments on the triple-axis spectrometer (TAS) DN1 installed at the Siloe reactor (CEA-Grenoble). The sample was fixed to the cold-tip of a orange type He cryostat with its $<001>$ axis vertical. In order to improve the energy resolution, we used the following configurations: Natural - PG(002) monochromator - 30' - sample - 30' - PG(002) analyzer - 30' D detector or Natural - PG(002) monochromator - 60' - sample - 60' - PG(002)analyzer - 30' D detector, where 30' and 60' indicate 30 and 60 minutes angular collimation. The inelastic scans were performed at fixed incident energies $E_i = 25$ meV and $E_i = 14$ meV. The resolution in energy was typically in the range $0.8-1.3$ meV. In order to determine the spin wave dispersions in MnS$_2$, we performed constant-$Q$ scans at the base temperature $T$ = 2.6 K of the He cryostat along several symmetry directions. 

The present data were collected from $\rm MnS_2$ single crystal in zero-field in the multi-domain state. We have, however, done some trial experiments on $\rm MnS_2$ to check whether we could generate a single-domain state by applying magnetic field and found that a magnetic field of 5 T was not enough for this purpose. We admit that the data collected from the multi-domain crystal may contain spurious signals from other domains but we assume that the intensity of such signals is relatively weak. The agreement between the experimental and the calculated dispersion, although quite good, is not perfect and we suspect that this is due to the contribution from other domains.

\section{Results}

Fig. \ref{scans} shows constant-$Q$ energy scans from MnS$_2$ at the base temperature $T$ = 2.6 K of He cryostat at different $Q$ values. The data have been fitted by Gaussian line shapes and the determined energies have been used for the dispersion of spin waves in MnS$_2$ and compared with the fitted dispersion by using Heisenberg model Hamiltonian shown in Eq. \ref{Hamiltonian}. 

It is important to note that owing to cubic symmetry of the lattice structure, multiple magnetic domains are present in the single crystal below $T_N$ which are related by wave vectors (1/2, 0, 0), (0, 1/2, 0) and (0, 0, 1/2). As a consequence, spin wave scattering intensity along $q_x$ and $q_y$ can have contributions from $q_z$ of other domain. This leads to small differences of measured spin wave energy along $q_x$ and $q_y$ (for same mode) despite planar $C_4$ crystal symmetry. This can be taken care by obtaining single $k$ domain with magnetic field or uniaxial stress \footnote{We have used $q$ to denote the wavevector corresponding to the axes defined in Fig.\ \ref{spin}, and it is used for theoretical calculation in next section. On the other hand, $Q$ is used to denote the wavevector in the experimental setup.}.

\section{Heisenberg model of spin wave dispersion in M${\rm n}$S$_2$}
In MnS$_2$, Mn spins are in high-spin ($S=5/2$) state and are arranged in fcc structure. The type-III AF order of Mn spins on the fcc lattice is shown in Fig. \ref{spin}. Alternating layers stacked along the $z$ direction (shown in solid and dashed lines) constitute two identical fcc sublattices with spins shown in red and blue. Within each fcc sublattice, the spins are aligned antiferromagnetically in the $xy$ plane as well as along the $z$ direction. Within each layer, the AF order of Mn spins yields two magnetic sublattices. For the spin wave analysis as discussed below, it will be convenient to distinguish between the interactions involving spins in the same layer and in different layers, as shown in Fig. \ref{spin}. Thus, we will consider the antiferromagnetic Heisenberg model on the fcc lattice:

\begin{equation}
\mathcal{H} = J \sum_{\langle ij \rangle} \bm{S}_i \cdot \bm{S}_j + J' \sum_{\langle ik \rangle} \bm{S}_i \cdot \bm{S}_k + J_d \sum_{\langle il \rangle} \bm{S}_i \cdot \bm{S}_l + J'_d \sum_{\langle im \rangle} \bm{S}_i \cdot \bm{S}_m - D \sum_{i} (\bm{S}_i^z)^2
\label{Hamiltonian}
\end{equation}
where $J$ and $J'$ are the exchange interactions between nearest-neighbor (NN) and next-nearest-neighbor (NNN) spins in the same layer, whereas $J_d$ and $J'_d$ similarly correspond to NN and NNN  interactions between spins in different layers (Fig. \ref{spin}). It should be noted that while $J,J',J'_d$ couple spins within the {\em same} fcc sublattice, $J_d$ couples spins in {\em different} fcc sublattices. The single-ion anisotropy term $D$ breaks spin-rotation symmetry and favors $z$-direction ordering in spin space (for $D>0$), resulting in spin-wave gap and stabilization of type-III AF order. This term may originate from coupling between spin and lattice degrees of freedom.

\section{Spin wave dispersion}
\label{SecV}

For the spin Hamiltonian in Eq. \ref{Hamiltonian}, we obtain four spin wave branches corresponding to the four layers shown in Fig. \ref{spin} for the type-III AF order. The derivation of the spin wave energy expressions is discussed in the Appendix. The spin wave dispersions in the $q_x-q_y$ plane ($q_z = 0$) are obtained as:

\begin{small}
\begin{eqnarray}
\omega^1_{\bf q} &=& 2JS(2+r_{\perp}) \;\; \Bigg[ \Big\{ 1 + \delta_{\rm sia} - \frac{2r_{\parallel}(1 - \gamma'_{\bf q})}{2+ r_{\perp}} + \frac{2r_d}{2 + r_{\perp}} \cos{q_x \over 2} \cos{q_y \over 2} \Big\}^2 - \Big\{\frac{2\gamma_{\bf q} + r_{\perp}}{2+r_{\perp}} + \frac{2r_d}{2 + r_{\perp}} \cos{q_x \over 2} \cos{q_y \over 2} \Big\}^2 \Bigg]^\frac{1}{2} \nonumber \\
\omega^2_{\bf q} &=& 2JS(2+r_{\perp}) \;\; \Bigg[ \Big\{ 1 + \delta_{\rm sia} - \frac{2r_{\parallel}(1 - \gamma'_{\bf q})}{2+ r_{\perp}} - \frac{2r_d}{2 + r_{\perp}} \cos{q_x \over 2} \cos{q_y \over 2} \Big\}^2 - \Big\{ \frac{2\gamma_{\bf q} + r_{\perp}}{2+r_{\perp}} - \frac{2r_d}{2 + r_{\perp}} \cos{q_x \over 2} \cos{q_y \over 2} \Big\}^2 \Bigg]^\frac{1}{2} \nonumber \\
\omega^3_{\bf q} &=& 2JS(2+r_{\perp}) \;\; \Bigg[ \Big\{ 1 + \delta_{\rm sia} - \frac{2r_{\parallel}(1 - \gamma'_{\bf q})} {2+ r_{\perp}} + \frac{2r_d}{2 + r_{\perp}} \sin{q_x \over 2} \sin{q_y \over 2} \Big\}^2 - \Big\{ \frac{2\gamma_{\bf q} - r_{\perp}}{2+r_{\perp}} - \frac{2r_d}{2 + r_{\perp}} \sin{q_x \over 2} \sin{q_y \over 2} \Big\}^2 \Bigg]^\frac{1}{2} \nonumber \\
\omega^4_{\bf q} &=& 2JS(2+r_{\perp}) \;\; \Bigg[ \Big\{ 1 + \delta_{\rm sia} - \frac{2r_{\parallel}(1 - \gamma'_{\bf q})} {2+ r_{\perp}} - \frac{2r_d}{2 + r_{\perp}} \sin{q_x \over 2} \sin{q_y \over 2} \Big\}^2 - \Big\{ \frac{2\gamma_{\bf q} - r_{\perp}}{2+r_{\perp}} + \frac{2r_d}{2 + r_{\perp}} \sin{q_x \over 2} \sin{q_y \over 2} \Big\}^2 \Bigg]^\frac{1}{2} \nonumber \\
\label{eq_sw1}
\end{eqnarray}
\end{small}
and along the $q_z$ direction ($q_x=q_y=0$):

\begin{small}
\begin{eqnarray}
\omega^1_{\bf q} &=& 2JS(2+r_{\perp}) \;\; \Bigg[ \Big\{ 1 + \delta_{\rm sia} - \frac{2r_{\parallel}(1 - \gamma'_{\bf q})} {2+ r_{\perp}} + \frac{2r_d}{2 + r_{\perp}} \cos{q_z \over 2} \Big\}^2 - \Big\{ \frac{2\gamma_{\bf q} + r_{\perp}}{2+r_{\perp}} + \frac{2r_d}{2 + r_{\perp}}  \cos{q_z \over 2} \Big\}^2 \Bigg]^\frac{1}{2} \nonumber \\
\omega^2_{\bf q} &=& 2JS(2+r_{\perp}) \;\; \Bigg[ \Big\{ 1 + \delta_{\rm sia} - \frac{2r_{\parallel}(1 - \gamma'_{\bf q})} {2+ r_{\perp}} - \frac{2r_d}{2 + r_{\perp}} \cos{q_z \over 2} \Big\}^2 - \Big\{ \frac{2\gamma_{\bf q} + r_{\perp}}{2+r_{\perp}} - \frac{2r_d}{2 + r_{\perp}}  \cos{q_z \over 2}  \Big\}^2 \Bigg]^\frac{1}{2} \nonumber \\
\omega^3_{\bf q} &=& 2JS(2+r_{\perp}) \;\; \Bigg[ \Big\{ 1 + \delta_{\rm sia} - \frac{2r_{\parallel}(1 - \gamma'_{\bf q})} {2+ r_{\perp}} + \frac{2r_d}{2 + r_{\perp}} \sin{q_z \over 2}  \Big\}^2 - \Big\{ \frac{2\gamma_{\bf q} - r_{\perp}}{2+r_{\perp}} - \frac{2r_d}{2 + r_{\perp}} \sin{q_z \over 2}  \Big\}^2 \Bigg]^\frac{1}{2} \nonumber \\
\omega^4_{\bf q} &=& 2JS(2+r_{\perp}) \;\; \Big[ \Big\{ 1 + \delta_{\rm sia} - \frac{2r_{\parallel}(1 - \gamma'_{\bf q}g)} {2+ r_{\perp}} - \frac{2r_d}{2 + r_{\perp}} \sin{q_z \over 2} \Big\}^2 - \Big\{ \frac{2\gamma_{\bf q} - r_{\perp}}{2+r_{\perp}} + \frac{2r_d}{2 + r_{\perp}}  \sin{q_z \over 2} \Big\}^2 \Bigg]^\frac{1}{2} . \nonumber \\
\label{eq_sw2}
\end{eqnarray}
\end{small}

Here $r_{\parallel} = J'/J$, $r_{\perp} = J'_d/J$, and $r_d = J_d/J$ denote the ratios of the exchange constants, $\gamma_{\bf q} = \frac{1}{2}(\cos q_x + \cos q_y)$ and $\gamma_{\bf q}' = \cos q_x \cos q_y$ are the Fourier transforms corresponding to NN and NNN lattice connectivity, and $\delta_{\rm sia} \approx D/2J$ represents the scaled single-ion anisotropy term. We will refer to the modes labelled (1,3) and (2,4) as the high-energy and low-energy branches, respectively, as explained below. In the ideal cubic case, $J=J_d$ and $J'=J'_d$, so that $r_{\parallel}=r_{\perp} \equiv r$ and $r_d = 1$. We have used $r_{\parallel} = r_{\perp} = r$ throughout this paper even when there is departure from the ideal cubic case.

It is instructive to first consider the spin wave energy expressions given above in the absence of any symmetry breaking $(D=0$). The spin wave energies $\omega^{2,4}_{\bf q}$ vanish exactly for the zone-boundary modes $q_x=q_y=\pi/2$ in the ideal cubic case $r_d = J_d / J \rightarrow 1$. Highlighting the intrinsic magnetic frustration in the fcc lattice, this zero spin wave energy for the low-energy branches arises from an exact cancellation between the positive energy cost (unhealing) associated with twisting of unfrustrated spins and the negative energy cost (healing) for frustrated spins. 

\begin{figure}
\includegraphics[width=70mm]{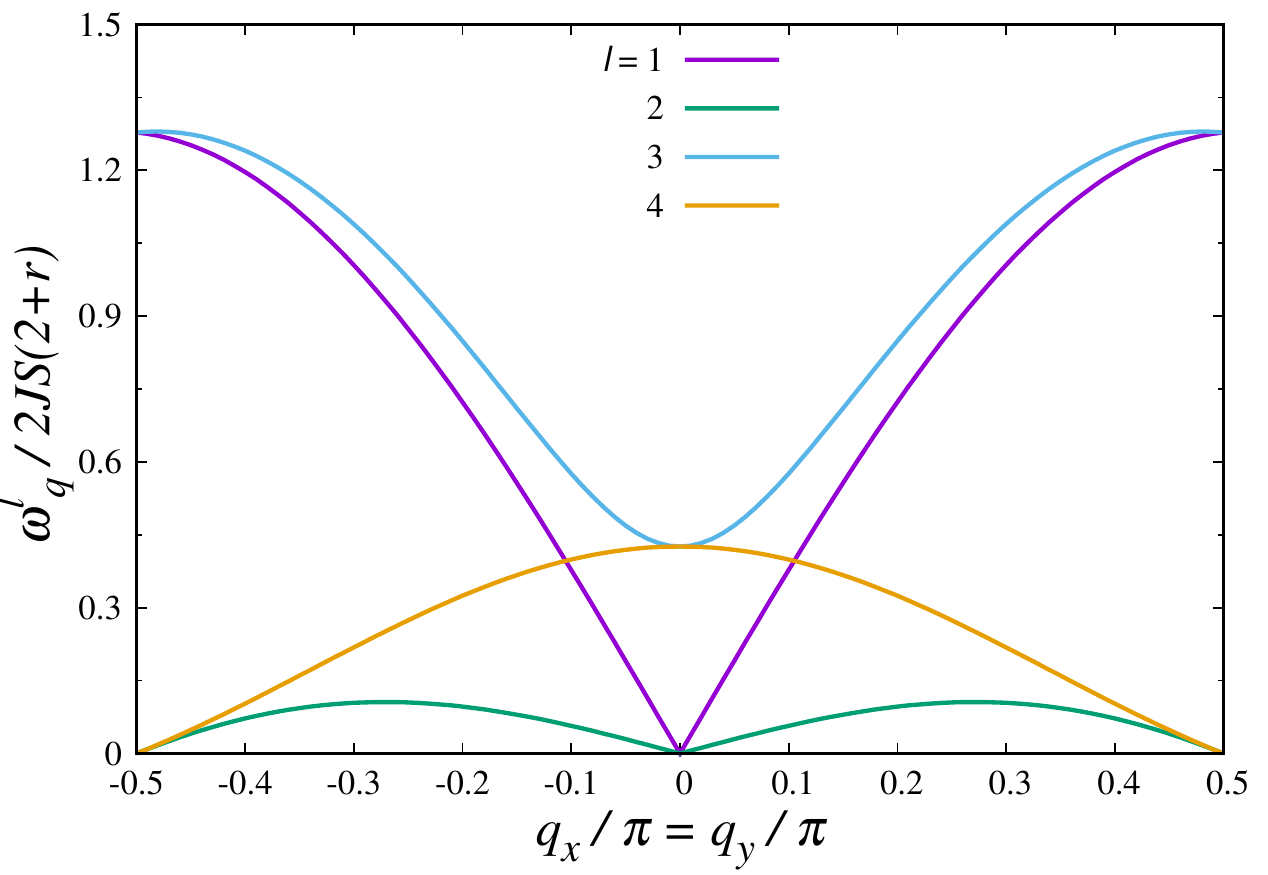}
\caption{Spin wave dispersions along the $q_x = q_y$ direction ($q_z$ = 0) for the spin rotationally symmetric ($D=0$) and ideal cubic ($r_d = 1$) case with $r = 0.1$, showing the high-energy ($ l =$ 1,3) and low-energy ($ l =$ 2,4) branches. The vanishing spin wave energy at $q_x = q_y = \pi/2$ for the low-energy branches highlights the strong intrinsic magnetic frustration in the fcc lattice.}
\label{sw_disp_zero}
\end{figure}

The full spin wave dispersions (Fig. \ref{sw_disp_zero}) show distinct separation into high-energy and low-energy branches. As discussed above, the low-energy branches reflect the strong intrinsic magnetic frustration in the fcc lattice, and the zone-boundary energy identically vanishes at $q_x = q_y = \pi/2$. The high-energy branches correspond to essentially independent magnetic excitations within each fcc sublattice, with only weak magnetic frustration due to the NNN interaction $J'$ between parallel spins. On the other hand, the low-energy branches reflect strong coupling between the two fcc sublattices. The ratio $r_d = J_d / J$ allows for this coupling to be tuned, and the low-energy branches become degenerate with the high-energy branches as $r_d \rightarrow 0$. 

\begin{figure}
\includegraphics[width=80mm]{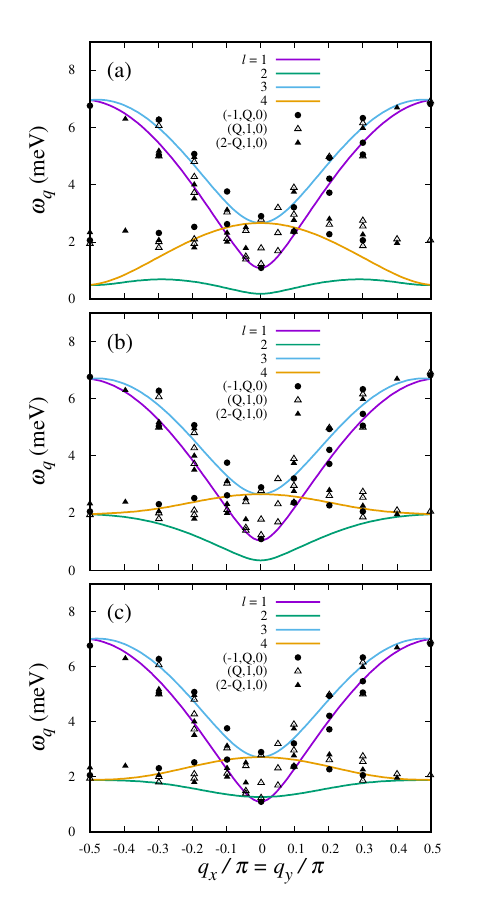}
\caption{Spin wave dispersions for modes $l=1-4$ along the $q_x = q_y$ direction calculated from Eq. \ref{eq_sw1} using different anisotropy models with: (a) only the single-ion anisotropy, (b) additional inter-layer exchange anisotropy, and (c) reduced exchange interaction for frustrated spins. The parameters used are mentioned in Sec.\ \ref{SecV}.}
\label{fig_sw}
\end{figure}

Before discussing the comparison with the experimental data, we will consider the effect of small deviation from the ideal cubic limit. Since bilinear exchange is separation dependent, anisotropic deformation which relatively decreases the exchange interaction between neighboring parallel (frustrated) spins as compared to antiparallel (unfrustrated) spins can result in significant magnetic energy gain, especially in a strongly frustrated system. Competition between structural distortion-induced elastic energy increase and magnetic energy lowering in the fcc lattice compound MnO with type-II AF order has been investigated in detail \cite{Lines1965b}. Octahedral rotations resulting in modified orbital overlaps can be another source of unequal exchange interactions. 

In view of the strong magnetic frustration in MnS$_2$, we will therefore allow for the possibility of slightly reduced exchange interactions between neighboring parallel spins resulting from anisotropic deformation of the cubic structure. It is unclear if this is related to the coupling between magnetic and lattice degrees of freedom in MnS$_2$, as indicated from the pseudo-tetragonal distortion observed below the magnetic ordering temperature in very high resolution synchrotron X-ray diffraction studies \cite{Kimber2015}.

For a given spin, while all four first neighbors in the same layer are unfrustrated (antiparallel), out of the 8 first neighbors in the adjacent layers, 4 are frustrated. We will therefore consider slightly reduced exchange interactions for these frustrated spins, denoted by factor $r_f$, in Eqs. \ref{eq_sw1} and \ref{eq_sw2} for the low-energy branches which involve strong inter-fcc spin-coupling effect. 

Fig. 4(a) shows the comparison of experimental data with calculated spin wave energies with only the single-ion anisotropy term included ($r_d = r_f =$ 1). Here $J'/ J =$ 0.12, $\delta_{\rm sia} =$ 0.01, and the scale $2JS =$ 2.60 meV. The high-energy branches are seen to fit well with the experimental data, including the spin-wave energy gap for the $l =$ 1 mode. However, the low-energy branches are significantly underestimated in this calculation. 

Fig. 4(b) shows that a better fit is obtained with the experimental spin wave energy ( $\sim$ 2 meV) of the low-energy branches at wave vector $\pi/2$ when the inter-layer exchange anisotropy term ($r_d = J_d / J = $0.85) is included. The other parameters are same as in (a).

Fig. 4(c) shows that further improvement in the fit is obtained for the lowest energy branch ($l =$ 2) on including the factor $r_f <$ 1 representing slightly reduced exchange interactions for the frustrated spins which significantly stabilizes the AF state. Here, $2JS$ = 2.65 meV, $r_f = $ 0.85, $r_d =$ 0.95, and other parameters are same as in (a). Therefore, we infer that description of experimental spin wave dispersion requires not only a small single-ion anisotropy, but slightly reduced exchange interaction for frustrated spins as well. For $S=5/2$ system, the values of spin interactions are $J \sim 0.5$ meV and $J' \sim 0.05$ meV, while the single-ion anisotropy $D \sim 0.01$ meV. It is important to note that the axes used for the calculation in Sec.\ \ref{SecV} are rotated by 45$^\circ$, and consequently $Q_x$ and $Q_y$ axis for experimental data corresponds to diagonal $q_x = q_y$ direction.

\section{Conclusion}

Study of magnetic excitations in $\rm MnS_2$ using inelastic neutron scattering measurements show distinct separation into high- and low-energy spin-wave branches, the latter highlighting the intrinsic magnetic frustration in the type-III ordered AF state on the fcc lattice. The spin wave spectra were found to be well described by the Heisenberg model with nearest- and next-nearest neighbor spin interactions and including a small single-ion anisotropy term. Comparison of the measured spin wave data with our theory shows an anomalous upward energy shift of the low-energy branch, which unambiguously indicates slightly reduced NN exchange interaction between frustrated (parallel) spins, and therefore suppressed magnetic frustration resulting from anisotropic deformation of the cubic structure.

\section{Acknowledgments}

S. G. acknowledges financial support from Deutsche Forschungsgemeinschaft through Collaborative Research Center SFB-1143. 

\appendix*

\section{Spin wave calculation}

\label{Appendix}

Spin wave analysis for the Heisenberg model is usually carried out using first the Holstein-Primakoff transformation which maps spins into bosons and then the Bogoliubov transformation to diagonalize the resulting linearized Hamiltonian. The large $8 \times 8$ matrix representation required for the type-III AF order on the fcc lattice renders this approach quite inconvenient. Instead, we will consider an alternative approach in terms of the equivalent itinerant electron picture, as discussed below. 

It is well know that the half-filled Hubbard model with $n^{\rm th}$ neighbor hopping term $t_n$ maps identically to the spin $S=1/2$ quantum Heisenberg antiferromagnet (QHAF) in the strong coupling limit, with effective spin interactions $J_n = 4t_n ^2 /U$. Furthermore, the spin wave energy expression obtained from the electronic approach within the random phase approximation (RPA) is identical to that obtained within the linear spin wave theory for the equivalent Heisenberg model. This has been analytically shown for several cases such as the planar, frustrated planar, and layered antiferromagnets \cite{Singh2001}. The key physical quantity which is evaluated for the AF state in the strong coupling limit is the bare particle-hole propagator $\chi^0({\bf q},\omega)$, in terms of which the spin wave propagator at the RPA level is given by $[\chi^{-+}({\bf q},\omega)] = [\chi^0({\bf q},\omega)] / ({\bf 1} - U[\chi^0({\bf q},\omega)])$, poles of which yield the spin wave energies. Essentially, since spin wave energies are associated with spin interaction energy corresponding to specific spin twisting modes, the effective spin couplings are inherently present in the spin wave propagator within the itinerant electron approach. 

The type-III AF order on the fcc lattice incorporates features of both the frustrated planar AF and the layered AF. The inter-fcc spin couplings $J_d$ between the two interpenetrating fcc sublattices is the only new feature, allowing for straightforward extension of the electronic approach to the fcc lattice antiferromagnet. Within the $8 \times 8$ matrix representation in the composite four-layer $\otimes$ two-magnetic-sublattice basis,\cite{Singh2017} this leads to the following expression for the ${\bf 1} - U[\chi^0({\bf q},\omega)]$ matrix in the strong coupling limit:

\noindent
\begin{equation*}
\begin{footnotesize}
\mathcal{C} \left[ \begin{array}{c c c c c c c c}
\mathcal{A} + \frac{\omega}{J(2+r_{\perp})} & \frac{2\gamma_{\bf q}}{2+r_{\perp}} & \frac{r_z}{2+r_{\perp}} c_{{\bf q}zxy} & \frac{r_z}{2+r_{\perp}} c_{{\bf q}z\overline{xy}} & 0 & \frac{r_{\perp}}{2+r_{\perp}} c_{{\bf q}z} & \frac{r_z}{2+r_{\perp}} c^*_{{\bf q}z\overline{xy}} & \frac{r_z}{2+r_{\perp}} c^*_{{\bf q}zxy} \\
& \mathcal{A} - \frac{\omega}{J(2+r_{\perp})} & \frac{r_z}{2+r_{\perp}} c_{{\bf q}z\overline{xy}} & \frac{r_z}{2+r_{\perp}} c_{{\bf q}zxy} & \frac{r_{\perp}}{2+r_{\perp}} c_{{\bf q}z} & 0 & \frac{r_z}{2+r_{\perp}} c^*_{{\bf q}zxy} & \frac{r_z}{2+r_{\perp}} c^*_{{\bf q}z\overline{xy}}\\
& &  \mathcal{A} + \frac{\omega}{J(2+r_{\perp})} & \frac{2\gamma_{\bf q}}{2+r_{\perp}} & \frac{r_z}{2+r_{\perp}}  c_{{\bf q}z\overline{xy}} & \frac{r_z}{2+r_{\perp}} c_{{\bf q}zxy} & 0 & \frac{r_{\perp}}{2+r_{\perp}} c_{{\bf q}z} \\
& & &  \mathcal{A} - \frac{\omega}{J(2+r_{\perp})} & \frac{r_z}{2+r_{\perp}} c_{{\bf q}zxy} & \frac{r_z}{2+r_{\perp}} c_{{\bf q}z\overline{xy}} & \frac{r_{\perp}}{2+r_{\perp}} c_{{\bf q}z} & 0 \\
& & & &   \mathcal{A} + \frac{\omega}{J(2+r_{\perp})} & \frac{2\gamma_{\bf q}}{2+r_{\perp}} & \frac{r_z}{2+r_{\perp}} c_{{\bf q}zxy} &  \frac{r_z}{2+r_{\perp}} c_{{\bf q}z\overline{xy}} \\
& & & & &  \mathcal{A} - \frac{\omega}{J(2+r_{\perp})} &  \frac{r_z}{2+r_{\perp}} c_{{\bf q}z\overline{xy}} &  \frac{r_z}{2+r_{\perp}} c_{{\bf q}zxy} \\
& & & & & & \mathcal{A} + \frac{\omega}{J(2+r_{\perp})} & \frac{2\gamma_{\bf q}}{2+r_{\perp}} \\
& & & & & & & \mathcal{A} - \frac{\omega}{J(2+r_{\perp})} \nonumber
\end{array} \right] 
\end{footnotesize}
\end{equation*}
where $\mathcal{C} = (2+r_{\perp})t^2/\Delta^2$, $\mathcal{A} = 1 + \delta_{\rm sia} - \frac{2r_{\parallel}}{(2+r_{\perp})} (1 - \gamma_{\bf q}')$, $\delta_{\rm sia} \approx D/2J$, $r_{\perp} = (t'_d/t)^2 = J'_d/J$, $r_{\parallel} = (t'/t)^2 = J'/J$, $r_d = (t_d/t)^2 = J_d/J$, $\gamma_{\bf q} = \frac{1}{2}(\cos q_x + \cos q_y)$, $\gamma_{\bf q}' = \cos q_x \cos q_y$, $c_{{\bf q}z} = \cos{q_z}$, $c_{{\bf q}zxy} = e^{ i q_z/2} \cos (\frac{q_x + q_y}{2})$, $c_{{\bf q}z\overline{xy}} = e^{ i q_z/2} \cos (\frac{q_x - q_y}{2})$.

The poles of the spin wave propagator are obtained by setting the eigenvalues of the matrix ${\bf 1} - U[\chi^0({\bf q},\omega)]$ to zero, which yields closed form expressions for the spin wave energies $\omega_{\bf q}$ as given in Eq. \ref{eq_sw1} and Eq. \ref{eq_sw2} for the four branches. Extension from the $S=1/2$ case to the general spin-$S$ case is incorporated by replacing $J$ by $2JS$.

The correctness of the matrix expression given above for ${\bf 1} - U[\chi^0({\bf q},\omega)]$ was confirmed from (i) proper interpolation to the known limiting cases corresponding to the frustrated planar antiferromagnet ($r_\perp = 0,r_d = 0$) and the layered antiferromagnet ($r_d = 0$), and (ii) comparison with the numerically obtained \cite{Singh2017} spin wave dispersion for the fcc lattice antiferromagnet ($r_d \ne 0$).

\bibliographystyle{apsrev4-1}

\bibliography{references}

\end{document}